\documentclass[a4paper,9pt]{article}

\usepackage{anysize}
\marginsize{40pt}{35pt}{8pt}{30pt}
\usepackage{placeins}
\usepackage[version=4]{mhchem}
\usepackage{siunitx}
\DeclareSIUnit\Molar{M}
\usepackage[utf8]{inputenc}
\usepackage{graphicx}
\usepackage{xcolor}
\usepackage{array}
\usepackage{multirow}
\usepackage{float}
\usepackage{amsmath}
\usepackage{amsfonts}
\usepackage{bbm}
\usepackage{microtype}
\usepackage{booktabs}

\usepackage{algorithm}
\usepackage{algpseudocode}



\definecolor{CCBlue}{RGB}{16, 38, 148}  
\usepackage[colorlinks=true,
            urlcolor=CCBlue,
            citecolor=CCBlue,
            linkcolor=CCBlue,
            bookmarks=true,
            breaklinks]{hyperref}

\usepackage[                
        backend=biber,      
        style=numeric-comp, 
        natbib=true,        
        hyperref=true,      
        alldates=year,      
	    uniquelist=false,   
        sorting=none,
	    isbn=false,
	    eprint=false,
	    doi=false,
        url=false,
	    maxcitenames=10,
        giveninits=true
]{biblatex}

\renewbibmacro{in:}{%
  \ifentrytype{article}{}{%
    \printtext{\bibstring{in}\intitlepunct}%
  }
}

\AtEveryBibitem{
    \clearfield{urlyear}
    \clearfield{urlmonth}
    \clearlist{language}
}

\usepackage{caption}
\begin{document}

\twocolumn[{\centering{\huge\textbf{Geometric developmental principles for the emergence of brain-like weighted and directed neuronal networks}\par}\vspace{3ex}
	{\Large 
	Aitor Morales-Gregorio\textsuperscript{1*}, Anno C. Kurth\textsuperscript{2}, and
    Karolína Korvasová\textsuperscript{1}\par}\vspace{2ex}
    \small 
    \textsuperscript{1}Faculty of Mathematics and Physics, Charles University, Prague, Czechia.  \\
    \textsuperscript{2}Hierarchical Neural Computation RIKEN ECL Research Unit, RIKEN Center for Brain Science, Wako, Japan.\\
    \textsuperscript{*}Corresponding author: \href{mailto:aitor.morales-gregorio@matfyz.cuni.cz}{aitor.morales-gregorio@matfyz.cuni.cz}\\
    \vspace{2ex}}

Brain networks exhibit remarkable structural properties, including high local clustering, short path lengths, and heavy-tailed weight and degree distributions. 
While these features are thought to enable efficient information processing with minimal wiring costs, the fundamental principles that generate such complex network architectures across species remain unclear. 
Here, we analyse single-neuron resolution connectomes across five species (\textit{C. Elegans}, \textit{Platynereis}, \textit{Drosophila M.}, zebrafish and mouse) to investigate the fundamental wiring principles underlying brain network formation. 
We show that distance-dependent connectivity alone produces small-world networks, but fails to generate heavy-tailed distributions. 
By incorporating weight-preferential attachment, which arises from spatial clustering of synapses along neurites, we reproduce heavy-tailed weight distributions while maintaining small-world topology. 
Adding degree-preferential attachment, linked to the extent of dendritic and axonal arborization, enables the generation of heavy-tailed degree distributions. 
Through systematic parameter exploration, we demonstrate that the combination of distance dependence, weight-preferential attachment, and degree-preferential attachment is sufficient to reproduce all characteristic properties of empirical brain networks.
Our results reveal that activity-independent geometric constraints during neural development can account for the conserved architectural principles observed across evolutionarily distant species, suggesting universal mechanisms governing neural circuit assembly.

\medbreak
\par\vspace{5ex}]

\section*{Main}

Brain networks are highly structured, combining strong local clustering, short path lengths, and heavy-tailed weight and degree distributions\supercite{Hilgetag2004,bullmore_economy_2012,SPORNS2004,Vertes2012,betzel_generative_2016}. These properties are believed to be the basis of efficient information processing while minimizing wiring costs\supercite{Kaiser2006}. Generative network models are used to study the neurodevelopmental processes that lead to the complex brain network properties observed across many species\supercite{Kaiser2004,Vertes2012,Jacob2015,lynn_heavy-tailed_2024}. 
However, the combination of biologically realistic and simple principles that collectively generate brain-like weighted and directed networks remains unknown.

The cost of creating and maintaining a connection to other neurons scales with physical distance in three-dimensional space\supercite{bullmore_economy_2012}. 
It thus appears reasonable that neuronal connectivity follows a distance-dependent rule\supercite{ercsey-ravasz_predictive_2013,perinelli_dependence_2019,Kurth2024}, with greater connection probability between nearby neurons. 
Several studies have focused on network generating models based on the distance between nodes, albeit with different definitions of distance.
Jost et al\supercite{jost_evolving_2002} investigated dynamically evolving networks by adding new nodes based on graph-theoretical distance (the number of indirect edges), resulting in scale-free degree distributions.
Kaiser and Hilgetag\supercite{Kaiser2004} showed that small-world networks can be achieved by adding edges to a network based on the probability derived from the physical distance between the network nodes.
However, distance dependence alone is not sufficient to generate all brain-like network properties noted above\supercite{Kaiser2004,Vertes2012,betzel_generative_2016}. 

In particular, heavy-tailed weight and degree distributions cannot be achieved from distance dependence alone\supercite{Kaiser2004,ercsey-ravasz_predictive_2013,Vertes2012,betzel_generative_2016}.
It is well known that preferential attachment, where connection probability scales with node degree or edge weight, produces networks with heavy-tailed degree\supercite{Barabasi1999} and weight distributions\supercite{lynn_heavy-tailed_2024}, respectively.

Another commonly used constraint is homophily: neurons with a high number of shared indirect connections are more likely to be directly connected\supercite{Jouve1998,van_albada_bringing_2020}.
Indeed, when distance dependence and homophily are combined, additional brain-like properties can be achieved\supercite{Vertes2012,betzel_generative_2016}, such as heavy-tailed degree distributions.
Betzel et al\supercite{betzel_generative_2016} explored several rules for generative models to match human connectivity from diffusion tensor imaging (DTI) and concluded that the physical distance between nodes and homophily are the most relevant constraints.
Homophily can also be derived from higher-order correlations of neuronal activity\supercite{lynn_heavy-tailed_2024}, which endows networks with clustered connectivity when used to determine connection probability, even without distance dependence.
However, homophily is not well explained by any known biological mechanism.
How does a neuron know that it shares many indirect connections with another not directly connected neuron?
What mechanism increases the probability that these neurons will actually connect?

\begin{figure*}[t]
    \centering
    \includegraphics[width=0.9\linewidth]{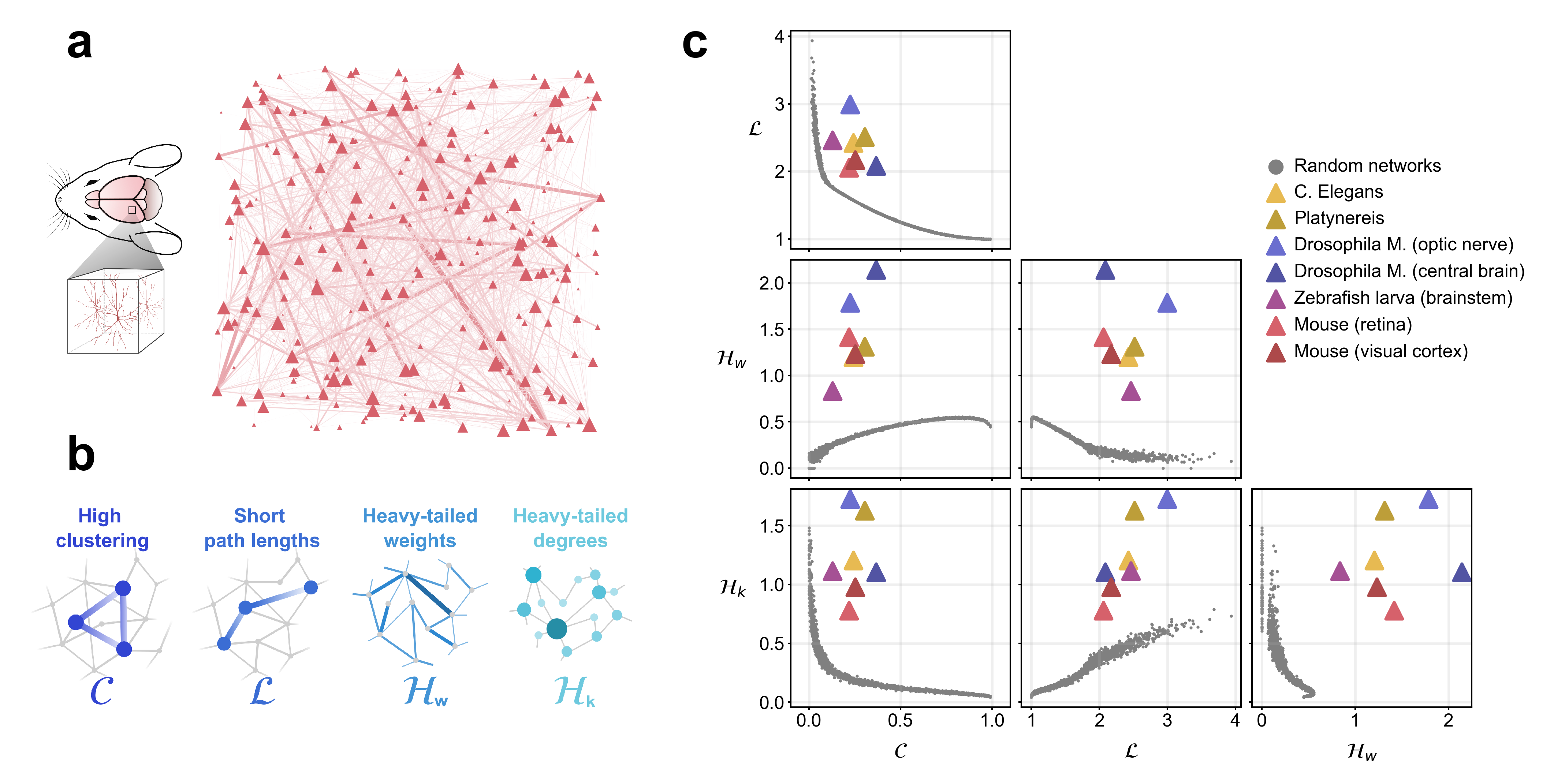}
    \caption{\small 
    \textbf{Empirical connectomes deviate from random networks.}
    \textbf{a)} Visualization of the empirical connectome from the mouse visual cortex\supercite{microns_2025}, neurons were placed at random locations on a lattice. The size of each node is determined by the degree, and the size of the edges by the connection weight.
    \textbf{b)} Schematic of properties found in all empirical connectomes.
    \textbf{c)} Pairplot of network properties: local clustering $\mathcal{C}$, average path length $\mathcal{L}$, weight heavy-tailedness $\mathcal{H}_w$, and degree heavy-tailedness $\mathcal{H}_k$. Properties of random networks (M=1840 sample networks of N=100 neurons per network, each sample has a different density $\rho \in [0.01, 0.95]$) and of empirical networks shown.
    }
    \label{fig:data}
\end{figure*}

Here, we leverage advances in brain connectomics to study the connectomes of five animal species at a single-neuron resolution\supercite{varshney_structural_2011,randel_neuronal_2014,takemura_visual_2013,scheffer_connectome_2020,Vishwanathan2024,helmstaedter_connectomic_2013,microns_2025}, from \textit{C. Elegans} to mouse. 
We investigate which biologically-realistic neuronal-wiring principles lead to complex brain-like networks.  
For each principle, we first provide direct biological evidence from the empirical connectomes and then use generative models to demonstrate which network properties can be achieved.
First, we confirm that distance dependence can lead to small-world networks, but not to heavy-tailed weight or degree distribution. 
Second, we show that including distance-dependent and weight-preferential attachment in combination can produce small-world networks with heavy-tailed weights, but not degrees. 
Finally, we show that the combination of distance-dependent, weight-preferential, and degree-preferential attachment can produce networks with all the properties of real brain networks.
Thus, we identified several fundamental principles of neuronal wiring, demonstrating that these three principles are both necessary and sufficient to reproduce brain network architecture across species.

\subsection*{Empirical connectomes deviate from random Poisson networks}

To characterize the properties of brain networks we study the single-neuron resolution empirical connectomes for \textit{C. Elegans}\supercite{varshney_structural_2011}, \textit{Platynereis} sensory motor circuit\supercite{randel_neuronal_2014}, \textit{Drosophila} optic medulla\supercite{takemura_visual_2013}, \textit{Drosophila} central brain\supercite{scheffer_connectome_2020}, zebrafish brainstem\supercite{Vishwanathan2024}, mouse retina\supercite{helmstaedter_connectomic_2013}, and mouse visual cortex\supercite{microns_2025}.
As an example, the connectivity from the mouse visual cortex\supercite{microns_2025} is shown in \autoref{fig:data}a.

It is common to compare empirical networks to some ``maximally random'' null model in order to determine their deviation from this randomness.
Since the empirical connectomes studied here are both weighted and directed, the unweighted Erd\H{o}s-Rényi (ER) model is not suitable.
Instead, we use a weighted version of the ER model, which we call the random Poisson model, see \nameref{method:model} for details.

The different brain networks share many properties (\autoref{fig:data}b) which deviate from random networks (\autoref{fig:data}c), see \nameref{method:props} for details. 
Our measurements are in agreement with previous reports\supercite{Hilgetag2004,bullmore_economy_2012,SPORNS2004,Vertes2012,betzel_generative_2016}, which concluded that brain networks tend to be small-world\supercite{Watts1998}, with high local clustering coefficient and small average shortest path length. 
At the same time, brain networks exhibit heavy-tailed weight distributions\supercite{lynn_heavy-tailed_2024,Cirunay2025, Piazza2025}, as well as heavy-tailed degree distributions\supercite{Cirunay2025,Reimann2025,Piazza2025}. 

Overall, brain networks across many species exhibit similar properties between them, but distinct from standard random networks. 
Similar properties across evolutionarily distant species suggest that similar developmental principles may be responsible for the complex structure of brain networks.

\FloatBarrier

\subsection*{Small-world networks emerge from distance dependence}

\begin{figure}[ht]
    \centering
    \includegraphics[width=\linewidth]{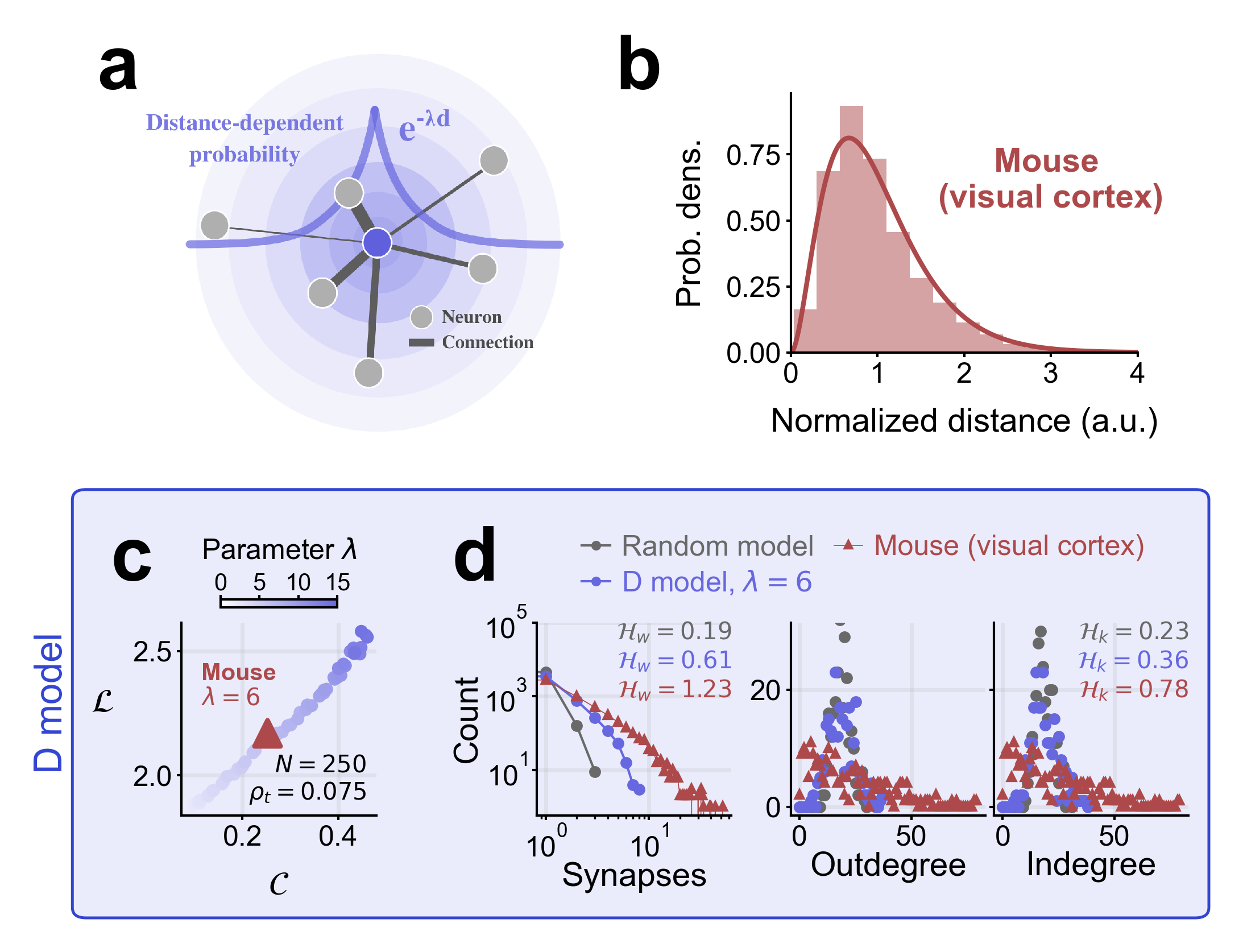}
    \caption{\small 
    \textbf{Distance dependence leads to high clustering and short path lengths, but not heavy-tailed degrees nor weights.}
    \textbf{a)} Schematic representation of distance-dependent connection probability with an exponential decay kernel.
    \textbf{b)} Distribution of synapses found at a given distance in the mouse visual cortex connectome. The empirical distribution (filled histogram) is well approximated by the combination of the expected distances within a bound sphere and an exponential decay kernel (bold line).
    \textbf{c)} Scatter plot of the clustering $\mathcal{C}$ and average path length $\mathcal{L}$ produced by the $D$ model with varying $\lambda$, especially tuned for mouse.
    \textbf{d)} Distributions of weight (left), outdegree (centre), and indegree (right) for the mouse visual cortex, random Poisson model and $D$ model.
    For the models, the distributions from the best fit parametrization are shown.
    Both models fail to match the distributions from the mouse connectome.
    }
    \label{fig:D_model}
\end{figure}

The connection probability between neurons has been suggested decrease exponentially with the distance between them\supercite{Markov2010,ercsey-ravasz_predictive_2013,Kurth2024} (\autoref{fig:D_model}a). 
Distance dependence is not surprising, as establishing long-range connections requires far more energy and resources than short-range ones\supercite{bullmore_economy_2012}. 
An important consideration is that the distance distribution is affected both by distance-dependent kernel and by the distribution of possible distances within the given bounded volume in which neurons reside\supercite{Kurth2024}. 
For neurons placed uniformly within a sphere, we confirm that the exponential kernel is the best fit to the data, see \nameref{suppl:distance_fit}.

To determine whether distance dependence is the fundamental principle behind the structure of the empirical connectomes, we introduce a simple neurodevelopmental model. Here, neurons are represented as points in space, neglecting their shape, neurite branching, and other features.
To initialize the network, $N$ neurons are placed randomly following an uniform distribution within a sphere of radius $R=1$, and the physical distances $S=[s_{ij}]_{i,j}$ between all model neurons are calculated.
In the following, the first index denotes the post-synaptic neuron, and the second index denotes the pre-synaptic neuron.
From these distances, we determine a probability distribution over all possible synaptic connections:
\begin{equation}
    \left(P_{D}\right)_{ij} = \frac{e^{-\lambda S_{ij}}}{\sum_{i,j} e^{-\lambda S_{ij}}}
    \label{eq:D_model}
\end{equation}
Here, $\lambda$ is the characteristic length of distance-dependent connectivity.
The weighted and directed adjacency matrix $A=[a_{ij}]_{i,j} \in \mathbb{R}^{N \times N}$ of the network is then constructed by iteratively adding new synapses to $A$. In each iteration, the newly added synapses are sampled from $P_{D}$. The process stops when an externally constrained target graph density $\rho_{t}$ is reached. In this way, we mimic neuronal wiring during development. Note that this process allows for multiple synapses between two neurons $i$ and $j$.
In the subsequent sections, this generation process is repeated while the probability distribution over the synaptic connections is gradually adapted to incorporate more principles of neuronal wiring.

The distance-dependent ($D$) model can generate networks that fit the clustering coefficient $\mathcal{C}$ and the average path length $\mathcal{L}$ from empirical connectomes (\autoref{fig:D_model}c). Although the networks obtained have heavier weight and degree distributions than random networks, they do not match empirical connectomes (\autoref{fig:D_model}d).

\FloatBarrier

\subsection*{Weight-preferential attachment}

\begin{figure}[ht]
    \centering
    \includegraphics[width=\linewidth]{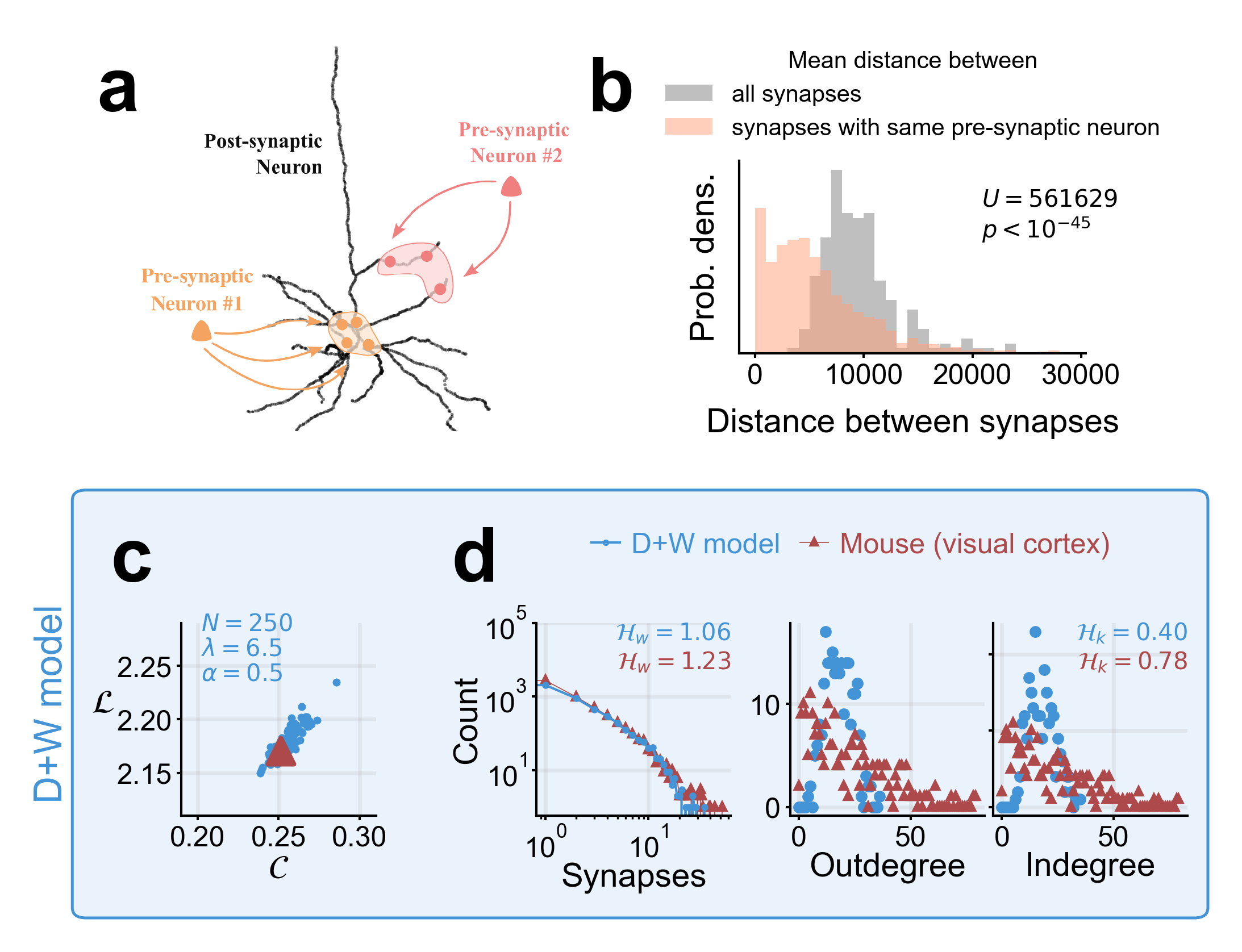}
    \caption{\small
    \textbf{Weight-preferential attachment principle.}
    \textbf{a)} Schematic of expected synapse distribution. Synapses originating from the same pre-synaptic neuron are expected to form with the same neurites, and therefore in close proximity to each other.
    \textbf{b)} Distribution of distance between synapses for all neurons in the mouse visual cortex\supercite{microns_2025}. The distance between synapses with the same pre-synaptic neuron is significantly smaller than the full distribution of distances between all synapses.
    \textbf{c)} Scatter plot of the clustering $\mathcal{C}$ and average path length $\mathcal{L}$ produced by the $D+W$ model. $N=100$ realizations with the same parameters.
    \textbf{d)} Distributions of weight (left), outdegree (centre), and indegree (right) for the mouse visual cortex and the best fit $D+W$ model.   
    }
    \label{fig:D+W_model}
\end{figure}

A well-established mechanism to achieve heavy-tailed weight distributions is preferential attachment\supercite{Barabasi1999}. 
That is, the probability of establishing a new synapse depends on the number of existing synapses between the two neurons. 
Biologically, this means that once a single synapse is established between two neurons, it is more likely that new synapses will form, likely because the neurite branches are already in close proximity to each other, a necessary condition for synapse formation as described by Peter's rule\supercite{Rees2017,van_albada_bringing_2020}. 
Within the dendritic tree, every synapse should be physically closer to other synapses originating from the same pre-synaptic neuron than to synapses originating from other neurons (\autoref{fig:D+W_model}a).
Indeed, this is exactly what we observe in the mouse visual cortex (\autoref{fig:D+W_model}b).
Therefore, the synaptic distribution suggests that weight-preferential attachment is a fundamental principle of synapse formation, governed mainly by the positions of the synapse-forming neurites.

To incorporate weight-preferential attachment ($W$) to our model, we use the number of existing synapses $A$.
The more synapses exist, the larger the neurite is expected to be and thus the probability of establishing new synapses increases. 
We consider the mixture distribution
\begin{equation}
    P_{D+W} = (1-\alpha) P_{D}+ \alpha \bar{A}, 
    \label{eq:D+W}
\end{equation}
where $\alpha \in [0, 1]$ and $\bar{A}_{ij} = A_{ij} / \sum_{i,j} A_{ij}$, i.e. the probability distribution of synaptic connections based on the realization of the network at a given iteration. $\alpha$ controls which fraction of the probability for a given synaptic connection between two neurons comes from the distance dependence and which from the weight-preferential ($W$) attachment.

The $D+W$ model generates networks with brain-like $\mathcal{C}$, $\mathcal{L}$ and $\mathcal{H}_w$ (best fit shown in \autoref{fig:D+W_model}c), but fail to show a heavy-tailed degree distribution (low $\mathcal{H}_k$).

\subsection*{Degree-preferential attachment}

\begin{figure}[ht]
    \centering
    \includegraphics[width=\linewidth]{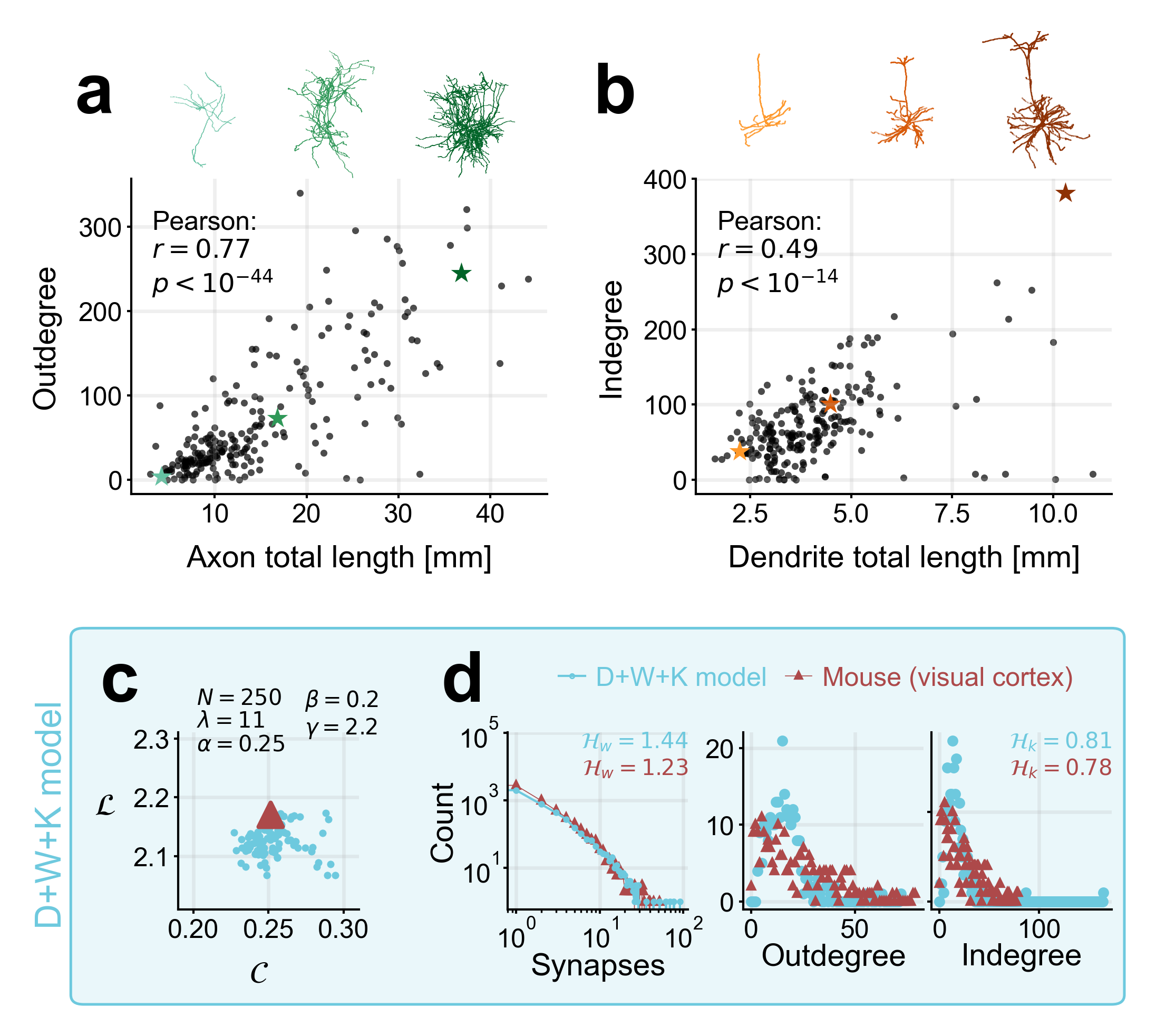}
    \caption{\small
    \textbf{Degree-preferential attachment principle.}
    \textbf{a)} Correlation between the total axonal length and outdegrees in the mouse visual cortex.
    \textbf{b)} Correlation between the total dendritic length and indegrees in the mouse visual cortex. Insets at the top of panels \textbf{a} and \textbf{b} are example morphologies from the data points indicated with stars.
    \textbf{c)} Scatter plot of the clustering $\mathcal{C}$ and average path length $\mathcal{L}$ produced by the $D+W+K$ model. $N=100$ realizations with the same parameters.
    \textbf{d)} Distributions of weight (left), outdegree (centre), and indegree (right) for the mouse visual cortex and the best fit $D+W+K$ model.   
    }
    \label{fig:D+W+K_model}
\end{figure}

We further extend the model to achieve heavy-tailed degree distributions. 
Following similar geometric arguments about neurite proximity, neurons with extensive dendritic branching are more likely to establish synapses with neurons with extensive axonal branching.
Furthermore, we predict that the in- and outdegree of the neurons is directly related to the extent of the pre- and post-synaptic neurite branching. 
We measured the neurite size, as the cumulative sum of the neurite lengths, of all axons and dendrites for the mouse visual cortex\supercite{microns_2025} by measuring the total length of the identified segments. 
There is a significant positive correlation between neurite length and the degrees (\autoref{fig:D+W+K_model}a,b).
Thus, we can use the in/out degrees as a proxy for the extent of neurite branching. 
Therefore, the degrees are related to the probability of new synapse formation because they capture whether the axons and dendrites are more likely to come in contact with each other due to larger branching.

To incorporate degree-preferential attachment ($K$) to our model, we consider the vectors $k_{\rm in}$ and $k_{\rm out}$ of in- and outdegrees of neurons in the current network iteration.
We define $K$ as the matrix that contains higher values for the edges that connect a neuron with a high outdegree to a neuron with a high indegree. $K$ is transformed into a probability distribution $\bar{K}$ by appropriate normalization.

The resulting probability distribution is a mixture of the three defined distributions; distance dependence, weight-preferential, and degree-preferential attachment:
\begin{equation}
    P_{D+W+K} = (1-\alpha-\beta) P_D + \alpha \bar{A} + \beta \bar{K}^{\gamma},
    \label{eq:D+W+K}
\end{equation}
with $\alpha, \beta \geq 0$ such that $\alpha + \beta \leq 1$. The parameter $\beta$ controls the fraction of the probability coming from the degrees, and $\gamma$ is an exponent controlling the nonlinear scaling of the probability. We impose $\alpha + \beta \leq 1$, such that the parameters $\alpha$ and $\beta$ determine which fraction of the probability will originate from each of the mechanisms: distance, weights, or degrees. 

The $D+W+K$ model can generate networks simultaneously matching the empirical connectomes for all the studied properties (\autoref{fig:D+W+K_model}c). 

Note that for degree-preferential attachment, we use a $\gamma$ exponent.
This exponent captures the nonlinear increase in probability as the volume of neurite branching increases.
If a linear scaling is assumed ($\gamma=1$), the $D+W+K$ model does not yield heavy-tailed degree distributions.
To achieve high $\mathcal{H}_k$, $\gamma > 1$ is required (\autoref{fig:D+W+K_model} bottom), suggesting that the size of neurite branching scales supralinearly with the degree.

\FloatBarrier

\subsection*{Parameter scans reveal capabilities and limitations of each model}

\begin{figure*}
    \centering
    \includegraphics[width=0.9\linewidth]{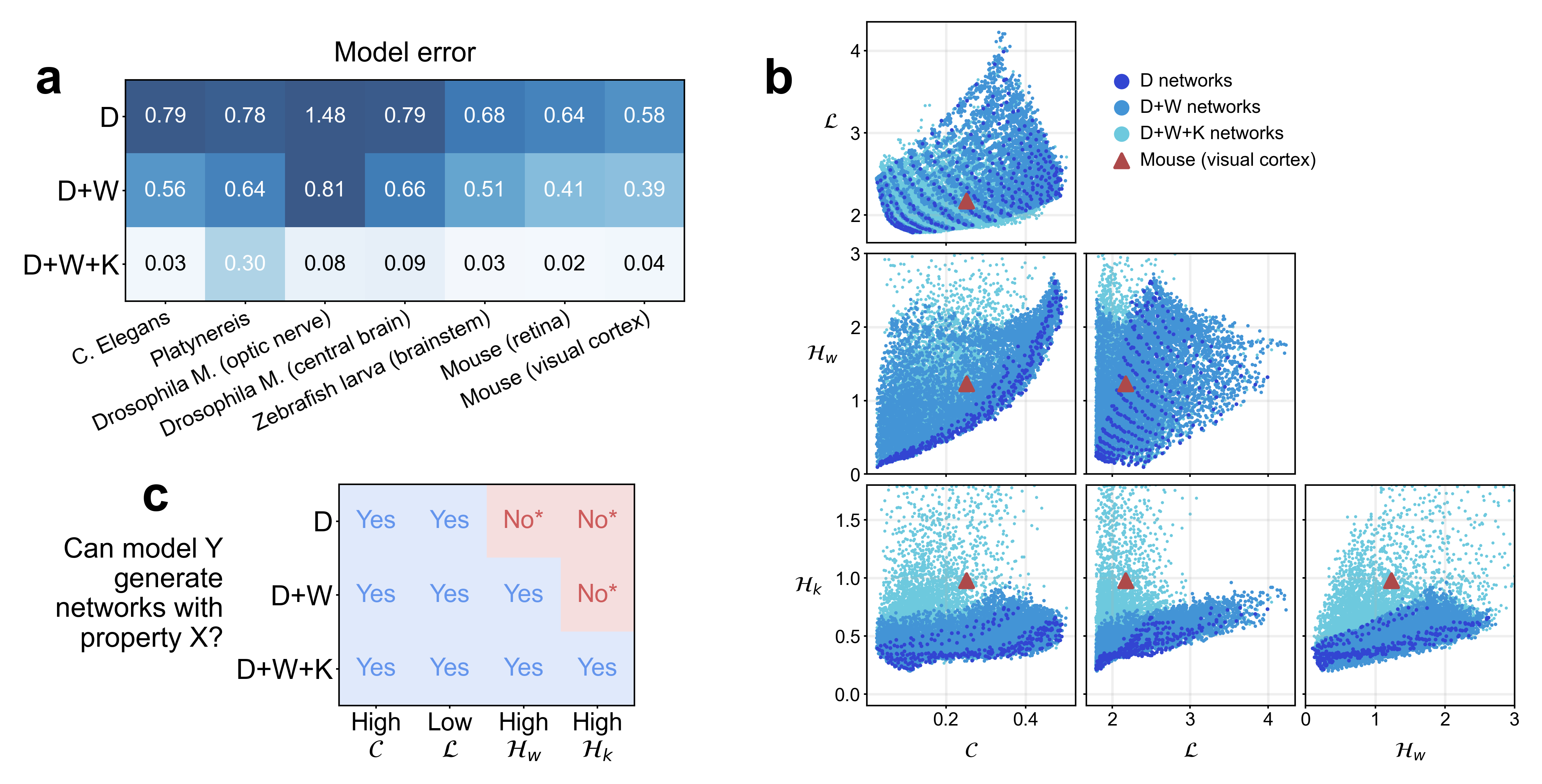}
    \caption{\small 
    \textbf{Parameter scans reveal capabilities and limitations of each model.}
    \textbf{a)} Error of best-fit model for each model and empirical connectome. Lower errors indicate a better match between the model and empirical networks.
    \textbf{b)} Results of parameter scan for the mouse visual cortex\supercite{microns_2025}. Each extension of the model increases the range of possible networks, but only $D+W+K$ can produce networks matching all the properties of the empirical network.
    \textbf{c)} Summary of whether each model can produce each of the four properties. ``No*'' dennotes that these properties cannot be achieved simultaneously with those marked as  ``Yes''. For example, the $D$ model can generate networks with brain-like $\mathcal{L}$ and $\mathcal{H}_w$, but then $\mathcal{C}$ is incorrect.
    }
    \label{fig:param_scans}
\end{figure*}

To better understand the capabilities and limitations of each generative principle, we performed extensive parameter scans.
The scans focused on identifying the optimal fit for each animal, by matching the number of neurons and target density to each case. 
In each case, we measured the error as the shortest Euclidean distance between the empirical and synthetic network properties in the feature space.
See \nameref{method:param_scans} for further details. 

We observe that in all cases the minimum error is found for the $D+W+K$ model (\autoref{fig:param_scans}a).
The Platynereis connectome has the largest error, an observation that might be explained by the relatively small sample size (N=59) when compared to the other networks.

Focusing on the mouse visual cortex (\autoref{fig:param_scans}b), the parameter scans show that some of the models are fundamentally limited. 
For example, the $D$ model cannot produce networks with brain-like $\mathcal{C}$ and $\mathcal{H}_w$; when one of them is correct the other will always be incorrect.
Equivalent plots to \autoref{fig:param_scans}b for all other animals are shown in \autoref{supfig:param_scan_others}. 
Incorporating further principles to the model expands the range of possible networks that can be generated (\autoref{fig:param_scans}b).
The $D+W+K$ model is the only model that can completely capture all brain-like properties of single-neuron networks.

\section*{Discussion}

In this work, we have identified three fundamental principles of neuronal wiring that are sufficient to generate brain-like networks across multiple species and organizational scales: distance dependence, weight-preferential attachment and degree-preferential attachment. First, we confirmed that distance-dependent connectivity alone produces small-world networks with high clustering and short path lengths\supercite{Kaiser2004,Watts1998}, consistent with wiring-cost minimization constraints\supercite{bullmore_economy_2012}. However, additional mechanisms are required to explain the heavy-tailed distributions observed in empirical connectomes\supercite{lynn_heavy-tailed_2024,Cirunay2025,Piazza2025} (\autoref{fig:data} \& \autoref{fig:D_model}).

Weight-preferential attachment emerges naturally from the spatial arrangement of neurites as predicted by Peter's rule\supercite{Rees2017,van_albada_bringing_2020}. Our finding that synapses from the same pre-synaptic neuron cluster spatially supports this geometric basis for preferential attachment (\autoref{fig:D+W_model}a,b). This mechanism generates heavy-tailed weight distributions while maintaining small-world properties (\autoref{fig:D+W_model}c,d). Degree-preferential attachment extends this principle by linking synaptic connectivity to neurite branching extent (\autoref{fig:D+W+K_model}). The observed correlation between neurite length and degree, combined with the supralinear scaling required in our model ($\gamma > 1$), suggests that larger dendritic and axonal arbours disproportionately increase the probability of synapse formation, potentially through increased overlap volumes or more branch points for potential synaptic contacts\supercite{Rees2017}.

For completion, we also studied the inter-area resolution networks measured from tract-tracing studies in mouse\supercite{Gamanut_2018}, marmoset\supercite{Majka2020} and macaque\supercite{Markov2012} cortex; see \nameref{suppl:area-level}.
Briefly, there were some key differences between single-neuron and inter-area resolution networks.
Particularly, inter-area networks had higher density and no heavy-tailed degree distribution.
Supplementary analysis showed that $D+W$ models were sufficient to capture most properties in the inter-area networks, although the simulated weight distributions did not fully match the empirical ones.
This analysis suggests different principles could be at play at the area and single-neuron levels.

Our findings at the single-neuron resolution align with established principles of neural development, where axon guidance and synapse formation proceed through sequential stages governed by molecular cues and intrinsic genetic programs\supercite{Sdhof2018,Sanes2020}.
During early development, axons navigate using molecular gradients and guidance cues\supercite{TessierLavigne1996} to reach target regions, establishing initial distance-dependent connectivity patterns.
Dendritic and axonal arbour growth is regulated by both intrinsic genetic programs and extrinsic signals\supercite{Jan2010,FerreiraCastro2020,Strner2022}, with more elaborate arbours providing greater opportunities for synapse formation, in agreement with our predictions that higher degree increases the connection probability (\autoref{fig:D+W+K_model}).
Synaptogenesis involves coordinated molecular events, including the recruitment of cell adhesion molecules, the assembly of pre- and post-synaptic scaffolding proteins, and cytoskeletal reorganization\supercite{Sdhof2008,Sdhof2018,Sdhof2021}.
Recent evidence suggests that these developmental processes follow similar principles across species\supercite{Reichert2008,Workman2013}, consistent with our cross-species findings.
Our model could be refined using cell-type specific connectivity patterns, mimicking the molecular gradients and biochemical cues in the brain.
Nevertheless, the geometric-principles studied here provide a sufficient explanation to the emergence of complex brain networks, suggesting that additional mechanisms are only required to fine tune the networks in later neurodevelopmental stages.

Neural activity alters the network structure through various plasticity mechanisms, yet it was not necessary in our model.
A recent generative model\supercite{lynn_heavy-tailed_2024} suggests that activity-dependent plasticity is required to achieve clustered networks with heavy-tailed weights.
However, the formation of a vast majority of synapses occurs normally in knockout mice with blocked neurotransmitter release\supercite{Verhage2000,Sigler2017,Sando2017}.
Moreover, neural activity is not required for the development of basic visual response behaviour in zebrafish\supercite{barabasi_functional_2024}.
Taken together, our results and the experimental evidence\supercite{Verhage2000,Sigler2017,Sando2017,barabasi_functional_2024} suggest that activity-dependent synaptic plasticity does not play a major role in shaping basic network properties during the early stages of neurodevelopment. 
Therefore, the activity-independent geometry-driven principles studied in this paper are a sufficient explanation for the emergence of brain-like properties in neuronal networks.

The consistency of our findings at the single-neuron level across five evolutionarily distant species (from \textit{C. elegans} to mouse), suggest that these principles reflect universal constraints on neural circuit assembly\supercite{SPORNS2004}. 
Future work incorporating more detailed morphological information\supercite{Reimann2025}, developmental time courses\supercite{Varier2011}, and molecular guidance cues could further refine these models and reveal species-specific variations in the relative contributions of each principle.

\section*{Methods}

\subsection*{Network properties}
\label{method:props}

Throughout this paper we focus on four network properties: average local clustering $\mathcal{C}$, average shortest path length $\mathcal{L}$, the heavy-tailedness of weight distribution $\mathcal{H}_w$,  and the heavy-tailedness of the (out)degree distribution $\mathcal{H}_k$. For all metrics, we exclude self connections, i.e. the diagonal of the adjacency matrix $A$. 

The local clustering coefficient for one node describes the fraction of neighbours of that node that are also connected between them. For simplicity, we use the unweighted and undirected definition of the clustering
\begin{equation}
    \mathcal{C} = \left< \frac{T_i}{k_i(k_i-1)} \right>_i
\end{equation}
where $T_i$ is the total number of triangles node $i$ is part of and $k_i$ is the degree. 
The networks are binarized and the directionality of the nodes is ignored for computing $\mathcal{C}$, because there is no agreed-upon definition for clustering in directed networks.

For the average shortest path length $\mathcal{L}$, the shortest path between all node pairs is computed from the unweighted and undirected network, so we assess the its width.
If the network is not fully connected, a length cannot be defined, so we manually set $\mathcal{L} = \infty$.

We define $\mathcal{H}_w$ and $\mathcal{H}_k$ as the Fano factor of the weight and degree distributions, respectively
\begin{equation}
    \mathcal{H}_w = \frac{\left< (A - \left< A \right>)^2 \right>}{\left< A \right>},\ \ \
    \mathcal{H}_k = \frac{\left< ({k_{\rm out} - \left< k_{\rm out} \right>)}^2 \right>}{\left< k_{\rm out} \right>}
\end{equation}
which is measured as the inverse product of the variance and the mean of the distribution.
The Fano factor yielded good approximations of the heavy-tailedness of the distribution. 
We also considered other metrics (e.g. Kurtosis, quantile-based metrics), but in practice they yielded very similar results to the simpler Fano factor. 

\subsection*{Empirical connectome preprocessing}
\label{method:data}

The empirical connectomes for \textit{C. Elegans}\supercite{varshney_structural_2011}, \textit{Platynereis} sensory motor circuit\supercite{randel_neuronal_2014}, \textit{Drosophila} optic medulla\supercite{takemura_visual_2013}, \\\textit{Drosophila} central brain\supercite{scheffer_connectome_2020}, zebrafish brainstem\supercite{Vishwanathan2024}, mouse retina\supercite{helmstaedter_connectomic_2013}, and mouse visual cortex\supercite{microns_2025} were measured from high-precision electron microscopy of in vitro tissue and the resulting connectomes were already published elsewhere. Here, we obtained the connectomes from the previously published data and performed a few preprocessing steps common to all connectomes.

For all data sets, we limit our analysis to manually proofread neurons. 
Due to the very small diameter of axons, it is common for automated reconstruction methods to miss large parts of the axonal branches. 
Manual proofreading has become the standard to solve this problem, where trained humans identify split branches and manually merge them. 
The lack of manually proofread neurons is the reason we did not include data from the human temporal cortex connectivity\supercite{ShapsonCoe2024}.

In some cases the data contained several disjoint networks, so we kept only the largest network component to avoid numerical issues and ensure accurate estimates of the network properties.
Furthermore, we exclude single neurons that have zero in- or outdegree, since they are likely only partially observed and lead to errors when measuring the network average path length.

\begin{table*}
    \centering
    \caption{Statistics of empirical connectomes for several animals.}
    \vspace{0.2cm}
    \begin{tabular}{r|c|ccccc}
    &   & Density & Clustering & Avg. path & Weight het. & Degree het. \\
    & N & $\rho$ & $\mathcal{C}$ & length $\mathcal{L}$ & $\mathcal{H}_w$ & $\mathcal{H}_k$ \\ \midrule
    C. Elegans                   & 274   & 0.040  &  0.24  &  2.42  &  0.43  &  1.21  \\
    Platynereis                  & 59    & 0.076  &  0.30  &  2.51  &  0.53  &  1.63  \\
    Drosophila M. optic nerve    & 836   & 0.017  &  0.22  &  2.99  &  0.39  &  1.73  \\
    Drosophila M. central brain  & 1000  & 0.087  &  0.37  &  2.08  &  0.54  &  1.11  \\
    Mouse retina                 & 984   & 0.038  &  0.22  &  2.06  &  0.44  &  0.78  \\
    Mouse visual cortex          & 250   & 0.086  &  0.25  &  2.17  &  0.46  &  0.98 
    \end{tabular}
    \label{tab:single_neuron}
\end{table*}

\subsection*{Network generating models}
\label{method:model}

Throughout this paper, we used \autoref{alg:model} to generate networks. Briefly, we initialize the algorithm with an $N \times N$ adjacency matrix $A$ where all entries are zero. We define the probability of adding a new synapse to each (potential) edge $P$ and use it to draw new edges at each iteration. The algorithm stops when a target graph density or maximum number of synapses is reached.
\begin{algorithm}
\caption{Network generating model pseudocode}
\label{alg:model}
\begin{algorithmic}[1]  
  \State Initialize an $N \times N$ empty network $A$
  \State Compute initial $P$
  \While{\parbox[t]{\linewidth}{%
         density($A$) $<$ target density \textbf{or}\\
         sum($A$) $<$ max synapses}
         }
    \If {$P$ depends on $A$}
        \State Update $P$ with current $A$
    \EndIf
    \State Sample edge $i,j$ according to $P$
    \State Add synapse to $A_{ij}$
  \EndWhile
\end{algorithmic}
\end{algorithm}

The main difference between the variants of our model is the probability distribution $P$ between the neurons:
\begin{itemize}
    \item In the random case, $P$ is the uniform distribution on the set of potential edges, the distribution of weights in this network approaches a Poisson distribution as $N\to\infty$. Thus, we call this model the \textbf{random Poisson model}.
    \item In the distance-dependent case, neurons are placed within a sphere of radius $R=1$, and $P$ depends on the physical distance between them (\autoref{eq:D_model}).
    \item In the weight preferential case, $P$ relies on the number of existing synapses between neurons (\autoref{eq:D+W}).
    \item In the degree preferential case, $P$ depends on the product of the pre-synaptic outdegrees and post-synaptic indegrees of the neurons (\autoref{eq:D+W+K}). 
\end{itemize}

In the weight and degree preferential cases, $P$ depends on the network itself; thus, at each iteration, the probability distribution $P$ must be updated. It is also necessary to initialize the network without preferential attachment; we usually do this with $N=1000$ synapses, either randomly or based on a distance-dependent probability. Without the initialization step, the network is likely to crumple, with all probability concentrated in a few synapses.

To accelerate the network generating process, we add multiple synapses per iteration: $m=100$ for low densities ($\rho < 0.1$) and $m=1000$ for higher densities ($\rho \geq 0.1$). 
Nevertheless, all the results were qualitatively equivalent to $m=1$ (not shown).

\subsection*{Parameter scans}
\label{method:param_scans}

We conduct parameter scans tailored to each animal connectome examined in this work. 
We match the number of neurons to each empirical connectome, setting a maximum of $N=1000$ neurons. 
Creating larger networks is possible, but it is prohibitively expensive for parameter scans.

For all cases, the target graph density is set to one of $20$ equally spaced values between $0.7 \rho$ and $1.3 \rho$, representing $\pm 30 \%$ of the empirical connectome density $\rho$.
The value ranges and number of samples for each parameter are shown in \autoref{tab:param_scans}.

\begin{table}[h]
    \centering
    \caption{Parameter ranges and number of samples used for parameter scans.}
    \vspace{0.2cm}
    \begin{tabular}{c|cc}
                                  & Range                 & Samples \\ \midrule
    Target density $\rho_t$  &  $[0.7\rho, 1.3\rho]$ & 20 \\
    $\lambda$                     &  $[3, 15]$            & 13 \\
    $\alpha$                      &  $[0, 0.95]$          & 20 \\
    $\beta$                       &  $[0, 0.95]$          & 20 \\
    $\gamma$                      &  $[0.6, 3.3]$         & 15   \\
    \end{tabular}
    \label{tab:param_scans}
\end{table}

\FloatBarrier

\section*{Data and code availability}
The processed data for all empirical connectomes, morphology, and results from parameter scans, as well as the Python code of network generating algorithms, parameter scans, data processing, and the code used to create all the figures are openly available at \url{doi.org/10.5281/zenodo.18185443}.

\section*{Competing interests}
The authors declare no competing interests.

\section*{Author contributions}
Conceptualization AMG, AK, KK;
Data curation AK, AMG;
Data preprocessing AMG;
Methodology AMG, AK, KK;
Software AMG; 
Formal Analysis AMG, AK;
Visualization AMG;
Writing - original draft AMG;
Writing - review \& editing AMG, AK, KK;
Supervision KK;
Resources AMG, KK;
Funding acquisition AMG, KK.

\section*{Acknowledgments}
We thank Jon Martinez Corral for his feedback, graphic design support, discussions, and patience; 
and Kayson Fakhar for useful discussions that helped contextualize this work.

This work received funding from the Programme Johannes Amos Comenius (OP JAK) under the project 'MSCA Fellowships CZ - UK3' (registration number \\CZ.02.01.01/00/22\_010/0008220); and from Charles University grant PRIMUS/24/MED/007.

\printbibliography


\clearpage
\setcounter{section}{0}
\renewcommand{\thesection}{S\arabic{section}}

\setcounter{figure}{0}
\renewcommand{\thefigure}{S\arabic{figure}}
\renewcommand{\theHfigure}{S\arabic{figure}}

\onecolumn

\section*{Supplement}

\subsection*{Determining the best kernel for distance-dependent connectivity}
\label{suppl:distance_fit}

The distance-dependent kernel is believed to be exponential\supercite{ercsey-ravasz_predictive_2013}. However, previous attempts to estimate the kernel's shape assumed the dependence on distance was the only variable. Yet, the geometry in which neurons are embedded also shapes the distribution of observable distances; longer distances are less likely to occur when sampling two random points within a given volume. For the 3-dimensional sphere, the probability of observing a specific distance is given by
\begin{equation}
    p_{sphere}(x) = 3\frac{x^2}{R^3} - \frac{9}{4}\frac{x^3}{R^4} + \frac{3}{16}\frac{x^5}{R^6}, \ \ \ \ 0 \leq x \leq 2R 
\end{equation}
where $x$ is the Euclidean distance between two uniformly sampled points, and $R$ is the sphere's radius. See section "2.6.3 Distance between two random points in a hypersphere"\supercite{Mathai1999} for a derivation of this probability density function.

Furthermore, the distance-dependent kernel transforms the distribution into a heavier-tailed one. In this section, we test three kernels: Gaussian, exponential, and Maxwell-Boltzmann. The Gaussian kernel is the simplest assumption, the exponential kernel is the best-known fit, and the Maxwell-Boltzmann distribution describes the probability of finding a particle's position influenced by Brownian motion (i.e., diffusion).

The kernels are:
\begin{equation*}
    p_{Gauss}(x) = \exp{\left(\frac{-x^2}{2\sigma^2}\right)}, \ \ \ \ \
    p_{exp}(x) = \exp{(-\lambda x)}, \ \ \ \ \
    p_{MB}(x) = \sqrt{\frac{2}{\pi}}\frac{x^2}{a^3} \exp{\left(\frac{-x^2}{2a^2}\right)}
\end{equation*}
where $\sigma$, $\lambda$, and $a$ are the corresponding parameters that control the spread of the distributions.

The observed distribution of distances is thus the product of the probability given by the geometry and the kernel $p(x) = \rm p_{geometry}(x) * p_{kernel}(x)$.

We test all three kernels against the distribution of distance between connected neurons in the mouse visual cortex\supercite{microns_2025}, macaque inter-area connectivity\supercite{Markov2012}, and marmoset inter-area connectivity\supercite{Majka2020}.
We assume a spherical geometry in all cases and calculate the best fit of each kernel using the minimum log-likelihood.

\begin{figure}[ht]
    \centering
    \includegraphics[width=0.7\linewidth]{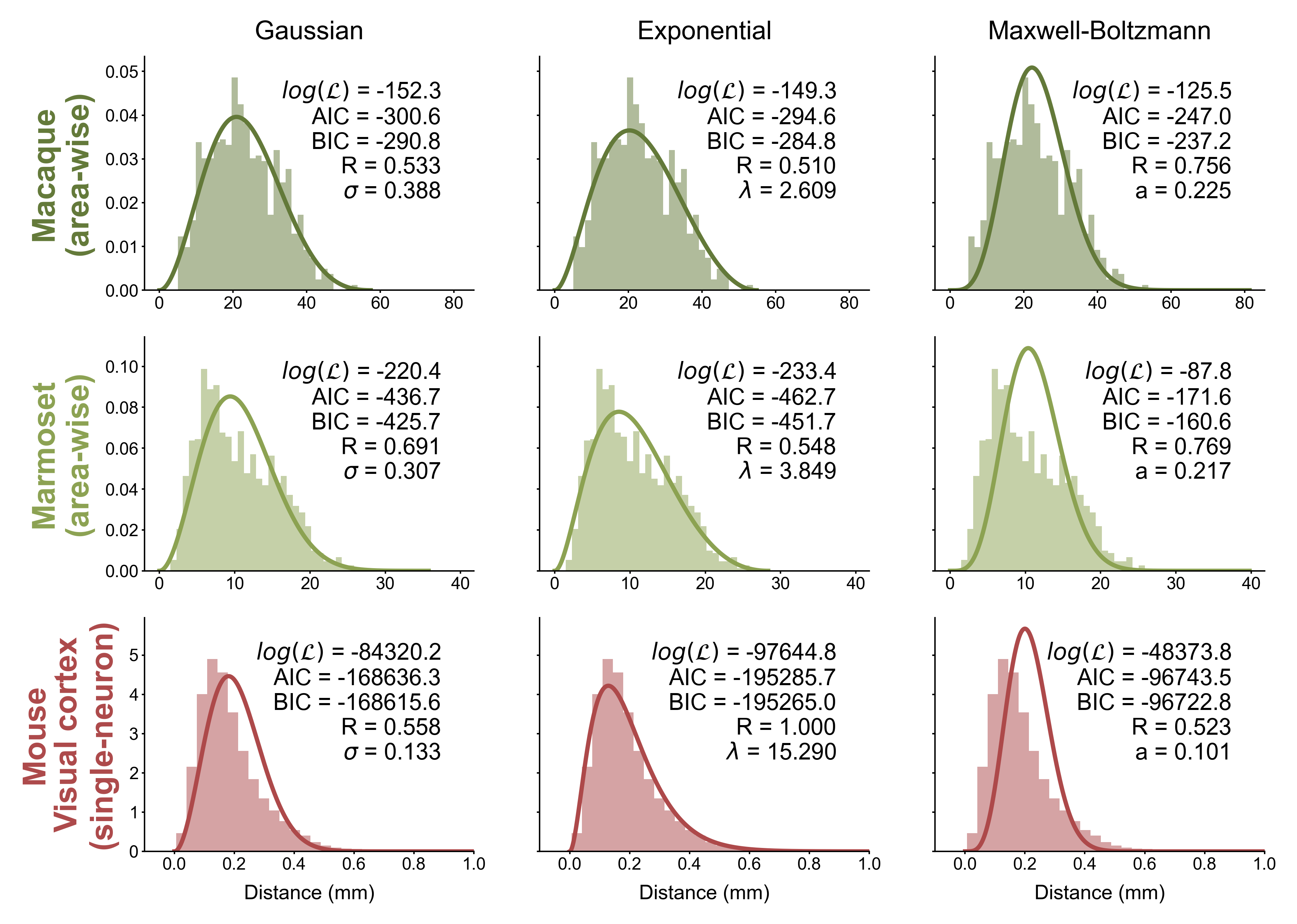}
    \caption{Fits of distance distribution approximations for mouse visual cortex\supercite{microns_2025}, macaque areas\supercite{Markov2012}, and marmoset areas\supercite{Majka2020}. 
    Filled histograms show the empirical distribution of distances between pre- and post-synaptic neurons from all connections in each data set. 
    The continuous lines show different fits with Gaussian, exponential, or Maxwell-Boltzmann distance-dependent kernels for each animal.}
    \label{supfig:dist_kernel}
\end{figure}

Our results in \autoref{supfig:dist_kernel} show that for marmoset and mouse, the exponential kernel produces the best fit, indicated by the highest negative log-likelihood, Akaike information criterion (AIC), and Bayes information criterion (BIC). 
For the macaque, the Gaussian kernel produces the best fit, closely followed by the exponential kernel.

\newpage
\subsection*{Parameter scans for all animals}
\label{suppl:param_scans}

\begin{figure}[ht]
    \centering
    \includegraphics[width=\linewidth]{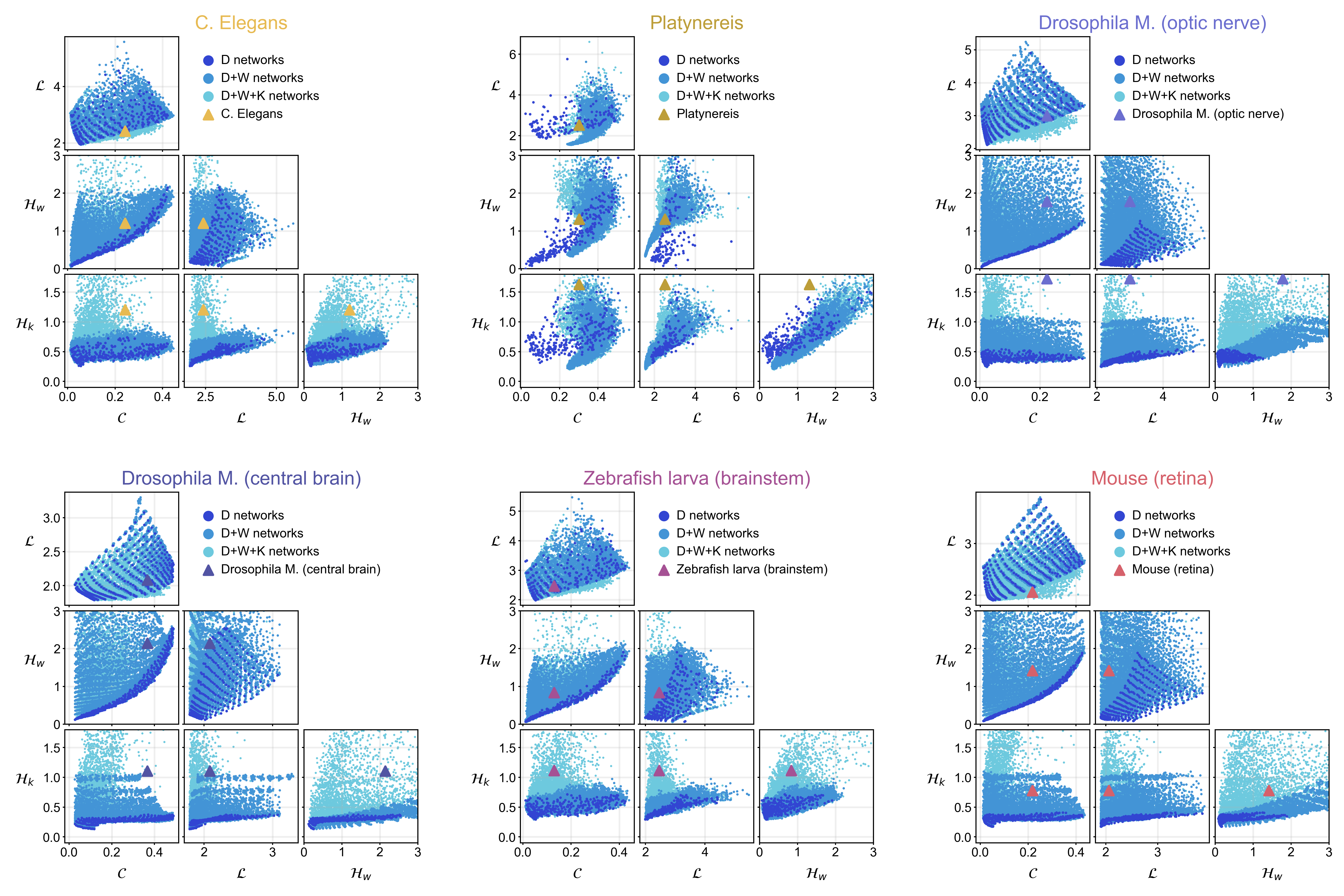}
    \caption{
        Parameter scans for all animals, akin to \autoref{fig:param_scans}b.   
    }
    \label{supfig:param_scan_others}
\end{figure}

\FloatBarrier
\subsection*{Inter-area brain networks}
\label{suppl:area-level}

All the networks studied in the main text of this work referred to connections between single neurons, measured using electron microscopy. However, many researchers focus on the connectivity between entire brain regions either from invasive tract-tracing experiments\supercite{Markov2012,Gamanut_2018,Majka2020} or non-invasive neuroimaging methods (EEG, fMRI, DTI etc)\supercite{betzel_generative_2016,Cirunay2025}.
To study whether distance-dependence, weight-preferential and degree-preferential attachment generalize from single-neuron networks to area-level networks, we also study the connectome at the inter-area resolution, derived from tract-tracing studies for mouse cortex\supercite{Gamanut_2018}, marmoset cortex\supercite{Majka2020}, and macaque cortex\supercite{Markov2012}.

The area-level tract-tracing connectomes fundamentally differ from the single-neuron resolution connectomes in several aspects.
First, the tract-tracing data does not consider the number of synapses between individual neurons, instead it quantifies how many pairs of neurons are connected between two brain regions. 
Second, the tract-tracing connectomes are not restricted to a single individual brain, instead they are the collation of many injections across many individuals (roughly between 10 and 30), with one or two injections per subject. 
Furthermore, tract-tracing is known to underestimate weaker connections and the studied connections are limited to the injection site which is usually only a small fraction of the entire brain area, increasing the undersampling problem.

\begin{table*}[h]
    \centering
    \caption{Statistics of empirical area-level connectomes for several animals.}
    \vspace{0.2cm}
    \begin{tabular}{r|c|ccccc}
    &   & Density & Clustering & Avg. path & Weight het. & Degree het. \\
    & N & $\rho$ & $\mathcal{C}$ & length $\mathcal{L}$ & $\mathcal{H}_w$ & $\mathcal{H}_k$ \\ \midrule
    Mouse cortex across areas    & 19    & 0.974  &  0.97  &  1.0   &  0.78  &  0.34  \\
    Marmoset cortex across areas & 55    & 0.624  &  0.75  &  1.25  &  0.81  &  0.20  \\
    Macaque cortex across areas  & 40    & 0.640  &  0.74  &  1.22  &  0.87  &  0.36
    \end{tabular}
    \label{tab:area_level}
\end{table*}

We measured all the properties of the area-level networks (\autoref{tab:area_level}). 
In contrast to the single-neuron connectomes (\autoref{fig:data}, \autoref{tab:single_neuron}), the area-level connectomes have fewer nodes (areas), much higher density, and low degree heavy-tailedness $\mathcal{H}_k$.

\begin{figure}[ht]
    \centering
    \includegraphics[width=\linewidth]{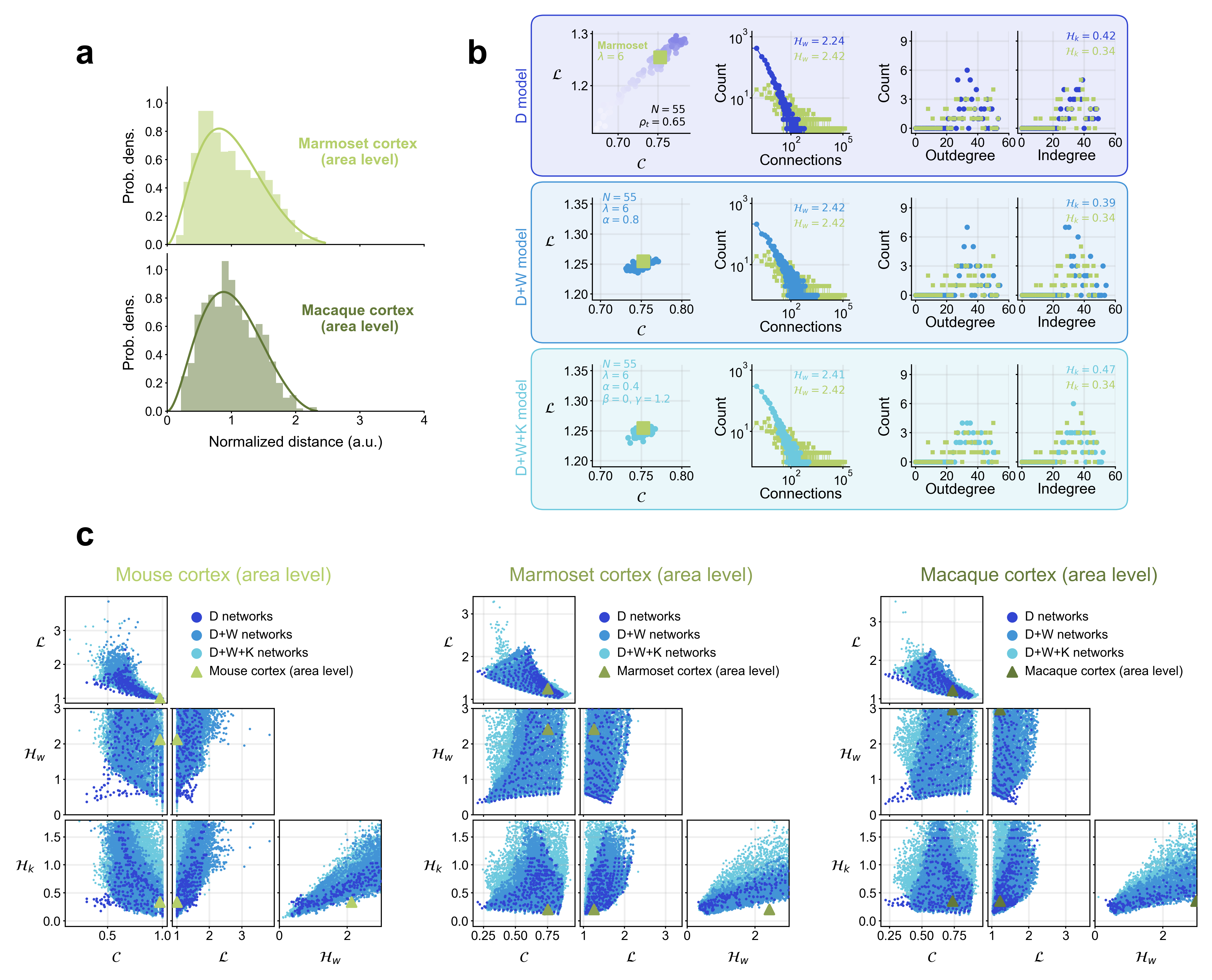}
    \caption{
    Our model does not generalize to inter-area brain networks.
    \textbf{a)} Distribution of connections found at a given distance in the marmoset and macaque area-level connectome. The empirical distribution (filled histogram) is well approximated by the combination of the expected distances within a bound sphere and an exponential decay kernel (bold line), as shown in \nameref{suppl:distance_fit}.
    \textbf{b)} Plot of $\mathcal{C}$ v. $\mathcal{L}$ (leftmost), weight (centre left), outdegree (centre right), and indegree (rightmost) distributions for the marmoset and the best fit simulated networks produced by the $D$, $D+W$, and $D+W+K$ models. 
    For the $D$ model multiple values of the parameter $\lambda$ are shown in the leftmost panel, whereas for the $D+W$ and $D+W+K$ models $N=100$ realizations with the same best-fit parameters are shown.
    All other panels show the distributions from the marmoset and the single best fit model realization.
    \textbf{c)} Parameter scans for the area-level connectomes, akin to \autoref{fig:param_scans}b.
    }
    \label{supfig:area_level}
\end{figure}

Distance dependence is observed in the area-level connectomes (\autoref{supfig:area_level}a), and the $D$ model can correctly approximate the small-world properties of the empirical connectomes (\autoref{supfig:area_level}b top). 
Incorporating the weight-preferential principle ($D+W$ model) leads to a better fit of the weight distribution (\autoref{supfig:area_level}b middle row). 
However, the weight distribution from the $D+W$ model does not perfectly match the empirical connectome, likely due to the bias in sampling from tract-tracing which underestimates weaker connections.
Since the area-level connectomes do not have heavy-tailed degree distributions, the $D+W+K$ model provides no improvement with respect to the $D+W$ model, in fact the best fit $D+W+K$ model parameter is $\beta=0$, i.e. no degree-preferential attachment at all.
These results suggest that different principles govern the connectivity at different scales.

\end{document}